\documentclass[conference]{IEEEtran}

\IEEEoverridecommandlockouts
\usepackage{cite}
\usepackage{amsmath,amssymb,amsfonts}
\usepackage{algorithmic}
\usepackage{graphicx}
\usepackage{textcomp}
\def\BibTeX{{\rm B\kern-.05em{\sc i\kern-.025em b}\kern-.08em
    T\kern-.1667em\lower.7ex\hbox{E}\kern-.125emX}}
\begin{document}

\title{Improve Blockchain Performance using Graph Data Structure and Parallel Mining\\
}

\author{\IEEEauthorblockN{Jia Kan}
\IEEEauthorblockA{\textit{Electrical and Electronic Engineering} \\
\textit{Department}\\
\textit{Xi'an Jiaotong Liverpool University}\\
Suzhou, China \\
Jia.Kan17@student.xjtlu.edu.cn}
\and
\IEEEauthorblockN{Shangzhe Chen}
\IEEEauthorblockA{\textit{Computer Science and Software } \\
\textit{Engineering Department}\\
\textit{Xi'an Jiaotong Liverpool University}\\
Suzhou, China \\
Shangzhe.Chen17@student.xjtlu.edu.cn}
\and
\IEEEauthorblockN{Xin Huang}
\IEEEauthorblockA{\textit{Computer Science and Software } \\
\textit{Engineering Department}\\
\textit{Xi'an Jiaotong Liverpool University}\\
Suzhou, China \\
Xin.Huang@xjtlu.edu.cn}
}



\maketitle

\begin{abstract}
Blockchain technology is ushering in another break-out year, the challenge of blockchain still remains to be solved. This paper analyzes the features of Bitcoin and Bitcoin-NG system based on blockchain, proposes an improved method of implementing blockchain systems by replacing the structure of the original chain with the graph data structure. It was named GraphChain. Each block represents a transaction and contains the balance status of the traders. Additionally, as everyone knows all the transactions in Bitcoin system will be baled by only one miner that will result in a lot of wasted effort, so another way to improve resource utilization is to change the original way to compete for miner to election and parallel mining. Researchers simulated blockchain with graph structure and parallel mining through python, and suggested the conceptual new graph model which can improve both capacity and performance.
\end{abstract}


\section{Introduction}

Blockchain is the backend of decentralized digital currencies one famous example of which is Bitcoin\cite{ref_article1}. Compared with traditional systems, blockchain has such features as being decentralized and immutable. According to the survey paper Bitcoin and Beyond\cite{ref_article5}, blockchain creatively combines existing contributions from decades of research to solve the foundational problem of passing the value over Internet.

Now Blockchain has become well-known and widespread. The field has gradually broadened, several limitations and disadvantages of the original design are showing up, one of the problems concerning the scalability of capacity and performance. The earliest famous example of blockchain network jam happened in a blockchain based game named CryptoKitties in 2017. Too many people were trying to purchase CryptoKitties over Ethereum\cite{ref_article7} network, soon Ethereum was exceeding its maximum processing capacity, all the transactions over Ethereum were delayed.

So far, Bitcoin\cite{ref_article1} and Ethereum\cite{ref_article7} have met the Transactions Per Second(TPS) bottom neck of performance. Both of them are looking for improvement solution. New protocol needs to be carried out as the new solution should address the previous blockchain's disadvantages. Commercial applications relay on the high performance blockchain. Most of public chain requires the ability to handle user requests from around the world. So the performance matters.

Limited open literature to date has reported, few studies have yielded in this field and the attempt to increase the performance was rarely made in academic world either. In contrast, many industrial blockchain products claim the performance as their products selling points, most of them approaching higher performance by tweaking blockchains parameters. There are only a few systematical methods that have been carried out, such as Bitcoin-NG\cite{ref_article2} (see below), Bitshares\cite{ref_article3}/EOS, IoTA and earlier DagCoin\cite{ref_article4}.

It is worth while mentioning that in paper Bitcoin-NG\cite{ref_article2}: A Scalable Blockchain Protocol, improvement was made by leader election. Nevertheless, only one leader was allowed to operate on the blockchain at one moment, which made the operation sequential, the capacity and performance still are a significant challenge. 

  Based on this idea, we proposed this method to increase the blockchain performance. For a high performance blockchain protocol which can increase the fundamental performance, data structure change is required. The chain data structure and Proof of Work algorithm are designed like a single-user operation system, only one user is allowed to touch the blockchain data. In our experiment, the parallel mining could visibly increase the TPS. Meanwhile, multi leaders could increase the network stability, to prevent the situation of network jam by no miners available.

Changing data structure from chain to graph can allow more than one leaders to mine parallelly. The graph data structure was named GraphChain in this experiment. GraphChain and parallel mining changed the blockchain to a multi-users operation system, more than one leaders was allowed to contribute their computing resources. It also benefited transaction confirm time.

This paper details the new protocol of a blockchain system. The key contribution of this work is the solution which provides a method to build a blockchain with practical performance. The new data structure plays the flexible role of framework and the new mining mechanism increases the overall performance. The outcome could be use scenario, such as a micro transactions system, blockchain with big data or a new generation banking system.

\section{Model and Goal}

It is well established that a blockchain system is consisted with N full nodes. Full node comes with the whole history transactions information, and the nodes are connected with reliable peer-to-peer network. Meanwhile there are M miners waiting for transaction requests and being ready to pack the transaction data into blockchain.

In the classic blockchain, the whole Miners are competing to solve a mathematics puzzle by iterating a number called nonce. The puzzle has a target with certain difficulty. When the puzzle is solved, a new block will be created. The recent transactions are packed into the new block. As will be described below, several problems has been discovered on this model:

1. Most of the power energy for calculation is wasted, as too much competition results in only one winner. Allowing one winner only means this miner acts like a central lock to limit the blockchains performance.

  2. The transactions need to wait for the next new block generation to be confirmed. It means transactions can not be finished in real time.

The construction of a better-performing blockchain system can be done by simply tweaking the blockchain parameters, for example increasing block size. However, from a technical survey's\cite{ref_article5} point, imply changing block size or time interval of block can not increase the performance and capacity all the way. There is a bottle neck from other aspects, such as ”the full node size limitation” or ”too much chain node forking”. Therefore a radically different design/model is needed to overcome this limitation. The researchers increased the performance and the capacity systemically and designed architecture that includes the data structure, mining mechanism and storage partition. See the concrete process of this experiment below:

1. Changed the current chain data structure into graph data structure. The changing of fundamental data structure made the following mechanism (see the 2rd below) improvement possible.

2. Changed the single miner to multiple miners for parallel mining. This improved the performance of blockchain and transactions would be handle in shortened time. The key contribution of this work was that it improved the performance of blockchain and transactions would be handle in shorten time.

3. Data shard. While the TPS was increased by parallel mining, data shard got the blockchain to store more transactions across different nodes. The full node would no longer exist as each node may store a different part of the whole transactions history.

This process of this experiment showed that the goal of changing the data structure to GraphChain was to break the central lock of traditional blockchain system, to get the scalable blockchain performance by parallel mining, increasing TPS and reducing transactions conformation time.

\section{Chain and Graph}

\begin{figure}[htbp]
\includegraphics[width=\linewidth]{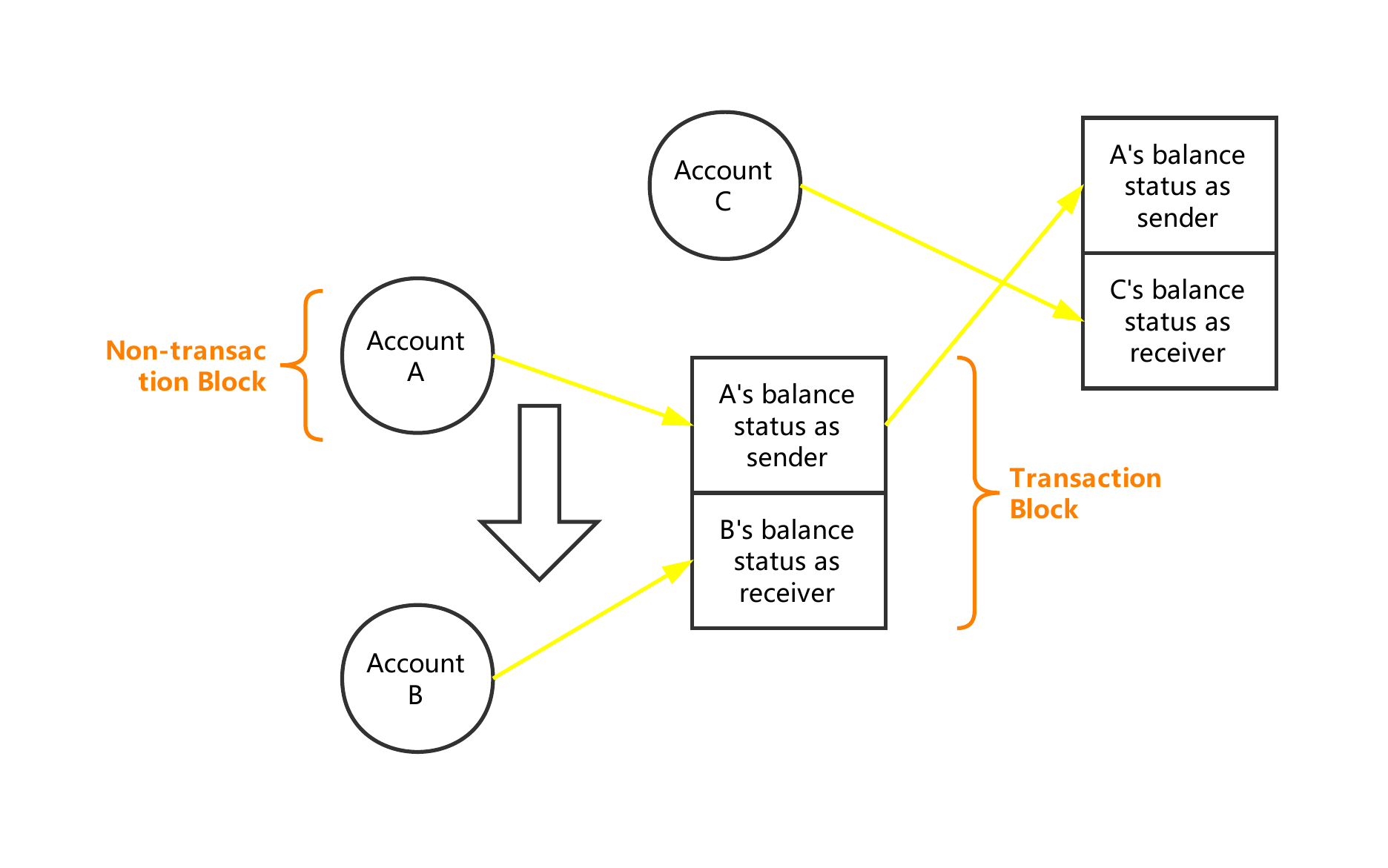}
\caption{The structure of GraphChain}
\label{fig1}
\end{figure}

As is well-known the chain data structure has been widely used in Bitcoin\cite{ref_article1} and other products. Directed acyclic graph(DAG) was used in IoTA and DagCoin\cite{ref_article4}. 

The GraphChain data structure idea was inspired from DAG. The main reason to change the data structure is to find a way to replace sequential operations with parallel operations. The data structure GraphChain is showed in Fig 1. GraphChain is a customized data structure, it's not a typicial DAG. The block usually has two inputs and two outputs, and the block represents the accounts' status after transaction. Another important point is that the inputs have the direction. If it's a sending account, it should connect to the input point above. Hence, the receiving account should be connected with the input point below.

Unlike chains, GraphChain don't need to pack the transactions into a single thread. With GraphChain, transactions can be added in parallel by different miners. 


\section{Parallel Mining}

The new mechanism is proposed to choose m (numbers of) leaders among M (as many as possible) miners world wide. In paper Bitcoin-NG\cite{ref_article2}, leader was introduced. Miners used to get rewarded by solving the puzzle and generating a new block. Now in parallel mining by solving the puzzle, miners have become the leader. The leader is on duty for the next period of time.

Unlike the traditional blockchain for example Bitcoin\cite{ref_article1}, the new mechanism requires the miner, who won the election, to stay online for its duty time. Sometimes accidents, like the power issue, or an Internet connection issue, happens unexpectedly. If the leader is offline, no one else can stay online to process the transactions instead of him. It makes the whole blockchain system to be blocked (transactions take too long time to be confirmed). In such cases, redundancy is required to keep the system running. It requires more than one leaders stand by to increase the stability of the system. The leaders work as a team. A set of rules is set up to get the miners cooperating. Below, there are specific details and elements:

\subsection{Miner to Leader}

On the basis of parallel mining, when a miner is elected as a leader, the leader serves his duty for a certain time. In a sliding time window, m (number of) miners are elected to leaders. According to Bitcoin\cite{ref_article1} system, the researchers set the election interval time to 10 minutes and the service time to 1 hour. It results that 6 leaders are online in a sliding time window. The leaders listen to the peer-to-peer network and wait for new transaction requests while they are on duty. Once the new transaction is broadcasting, leaders pack it into GraphChain. As the delay of communication between leaders, there are chances that same transaction is packed into the GraphChain multiple times by different leaders. Only m (number of) leaders are working in parallel, resulting in maximum of m (number of) forks of the GraphChain.

\subsection{Accounts}

An account usually starts with zero balance. Account balance changes because of transactions. At least two accounts are involved within a transaction. The account block is a non-transaction block, usually only having one output(Fig 1). The transaction block represents two account's balance status.

\subsection{Issue and Revoke Assets}

The block issuing or revoking assets requires special signature from certain private key, or more than one keys. Any non-transaction block with assets above zero should be verified carefully in the mechanism design.

\subsection{Transactions}

Any transaction should happen between at least two ac- counts. The senders balance must be above zero. As shown in Fig 1, each transaction block usually has two inputs and two outputs. A transaction block stands for two accounts current balances, one of them being a sender account and the other a receiver account.

\subsection{Scalability}

The scalability of Bitcoin\cite{ref_article1} and related system is limited. The reasons include block size, the overall data amount, the block propagation speed and miner's capacity. In Bitcoin\cite{ref_article1}'s design, one miner is selected to pack the transactions from the past 10 minutes. The rule acts like a central lock, preventing other miners from contribution.

Parallel mining enables independent transactions to happen at the same time. Let's take a look at two types of scenarios in real life: 1. the parents give the kid some pocket money 2. the shop accepts money for goods sold. The transaction of pocket money example happens in every family. They are independent, lots of parents can transfer money in parallel. The second type of transactions is made in sequence, the commerce accepting each transaction will change its own balance. Both types of transactions are very common, sometimes they're mixed. Parallel mining will speed up transactions processing by allowing more leaders to play the mining task.

\subsection{Election for the Leaders}

In GraphChain and Bitcoin-NG's design, leaders play the role of Bitcoin\cite{ref_article1}'s miner. In Bitcoin, miner who finds the new block has the privilege to pack the transactions from the past 10 minutes into that new block. In GraphChain system, leaders are elected to pack the transactions for the certain time. It comes with a problem: the Bitcoin's miner finishes his task once the mining is over, the GraphChain and Bitcoin-NG's leaders need continue to work once the election is done. Any leader may go offline any time due to exception, it unable to ensure availability unless adding redundancy. In GraphChain or Bitcoin-NG, if there is no leader online, the whole system will be jammed， that's why there should be more than one leaders online as the replicas. Leaders work together while they are competing. The amount of leaders depends on system performance requirement.

The leader is elected in the similar way that Bitcoin miner finds the new block. The method is PoW based. A miner who wants to be an leader, the world wide competition is required. The difference is that Bitcoin rewards the new block to the winner miner, GraphChain grants the qualify to the winner miner. The miner became a leader will work with other leaders for a while to pack the user transactions into GraphChain.

\subsection{Append Transaction}

The original PoW algorithm sets a difficult target. Whoever first calculates the puzzle will broadcast the result and create the next block. In this case, there are unlimited competitors and only one of them can win.

In parallel mining scenario, limited number of leaders try to solve much simpler puzzle. If a leader retries certain times and still can not find the result meeting the target, this leader will give up. Because the puzzle is simple enough, it's assumed that at least one leader should be able to get the puzzle answer and create the next transaction block.

The result is that several blocks may be created by different leaders for one transaction. The block whose creator spends less retries will be the lucky new block. The simple modification of PoW algorithm is called Proof of Luck(PoL).

The leaders append the transaction to the GraphChain data structure. As there are a limited number of leaders, PoL algorithm plays an efficiency way. Any leader can choose from any unpacked transaction to append into GraphChain, and different orders are fine.

For efficiency, algorithm can be applied that different leaders rank transactions by different orders, to make sure all the transactions should be appended into data structure for at least once. This is a kind of redundancy that the computation is verified.

\section{Data sharding}

The chain data structure is hard to be 'sharded' because the blockchain can not be split by account.

The existing solution suggests keeping the certain period of times blockchain by snapshot. The accounts balance list also need to keep stand-alone. Dropping part of the blockchain data will cause the difficulty in data verification. 

Also, the term 'sharding' here is different from Ethereum\cite{ref_article7}'s 'sharding', which is described at Github project page https://github.com/ethereum/sharding/blob/develop/docs/doc.md. Ethereum's sharding is to increase the capacity with a two-layer design. In GraphChain, the sharding can be done by splitting data by user.

As the high TPS's requirement for the production environment, the overall size of data will be increased very fast. Assume there are 2000 TPS, about 24G data will be generated per day. Less PC will meet the condition to act as the full clone node.

Data shard on GraphChain divides the data by accounts. In general a user account internally is represented as hash. A node can be simply designed to keep all the transaction information related the accounts whose hash starts with 'a' or 'b'. By this configuration, the huge data can be kept separately by many nodes with redundancy.

In GraphChain, each block represent the history of two accounts’ states. Verifying the correctness of transactions can split into two steps: verify all the transactions history related to one user account and repeat the same action on all the user accounts. When the history data is big enough, it will be a nice choice to have the parallel verification on multi nodes.

\section{Metrics}

To prove the performance improvement proposed, single leader and multi leaders performances need measuring.

The first experiment(Fig 2) assumed 4000 transactions were waiting to be packed, and performances were measured by the different numbers of leaders. 4000 transactions under waiting is actually a big jamming case. In practical the new transactions should be processed as soon as possible.

In this experiment, the leaders are free to choose any transactions to write to GraphChain. If network delay, a leader may try to pack a transaction which is already done by other leaders. According to the longest chain rule, in the case of two chains with same length, PoL algorithm is used to decide which fork to go.

It's also found that the overall performance would be going down while more transactions are appended into GraphChain. The reason is when the chain length increases, the database need to lookup the chain to verify if the transaction exists already in the blocks. This issue is solved by adding cache. After applying this technique we can observe quite average performance in Fig 2.


The second experiment(Fig 3) measured the stability of the blockchain network. If all the leaders were offline, the network was in jam as no transaction could be processed. Researchers used Simple Gilbert model\cite{ref_article6} method to model the network status. It was assumed that all the leaders each had chance p to go offline and 1-p chance to get back online. The stability for solo leader and multi leaders was measured.

\section{Experimental Setup}

\begin{figure}[htbp]
\includegraphics[width=\linewidth]{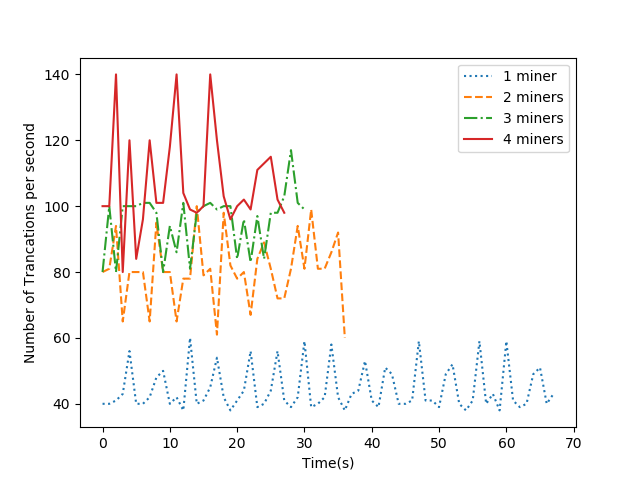}
\caption{The performance of parallel mining by adding miners}
\label{fig2}
\end{figure}

The experimental source code is listed on Github https://github.com/ProofOfLuck/graph

MySQL database was used to store the graph, leaders and transactions data. Its assumed that the peer-to-peer network transportation was reliable. The experiment didn't implement that part. In Bitcoin, data is stored in different full-nodes, and be replicated (copy from each others). This part is already be proven to be feasible. In GraphChain, this part is following the Bitcoin. Hence, the data is stored in centralized MySQL database would not affect performance measure.

Transactions are inserted into the databases table directly. Below, there are two major steps and details for each script files:

\subsection{Simulate the Election}

The miners first become leaders through election. In real cases, difficulty and competition was required to the election. In this simulation code, miners will become a leader by solving simple puzzles (iteration hash with some fixed target).

\subsection{Simulate the Mining}

The leaders observe the transactions table of database to see if new transaction turns up. Once new transaction appears, the free leaders query in the GraphChain to see if other leaders have finished packing the transaction into block already. If not, packing will be started by the current leader immediately. Check the code below (the experimental source code in Github repositories https://github.com/ProofOfLuck/graph):

election.py elects the miner to become a leader by solving simple hash based puzzles. leader.py plays the leaders role, packing transaction into block. leader auto.py combines the election and leaders function, keep playing leaders role and re-election when the leader period expires.

send.py is a script for user to send balance to another user.
send\_auto.py simulates auto transaction between a group of users.

wallet\_new.py generates a private key. The private key is the identity, could be used for both miner or normal user. To keep it simple, the public key of the identity will be used as wallet address.

The first experiment(Fig 2) looked for the proof that how much performance the blockchain could achieve. In Fig 2, different number of miners are simulated in mining tasks from the same amount of transactions. It's observed that more transactions were processed by the system, as the number of miners working on the GraphChain grew. This simulation was finished on one computer. The simulation code had been adjusted to slow down intentionally to prevent over using CPU.

The second experiment(Fig 3) was to simulate that miners each had the chance to go offline and get back online. Simple Gilbert model\cite{ref_article6} was used in this simulation. From the result, it's observed that with 5 miners working together(p set to 0.1 and q set to 0.5) there was little down time.

\begin{figure}[htbp]
\includegraphics[width=\linewidth]{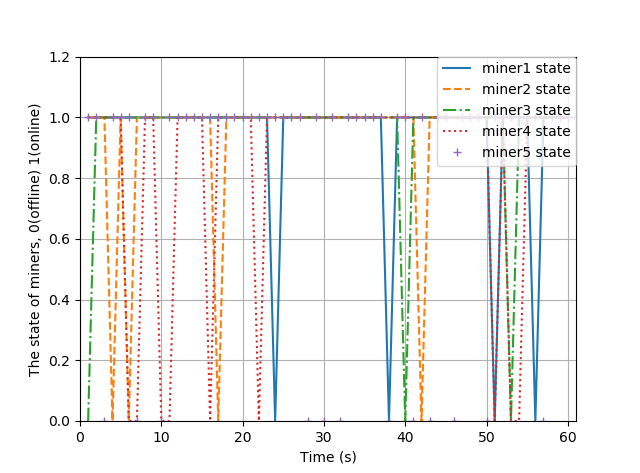}
\caption{The availability of multi miners}
\label{fig3}
\end{figure}

\section{Evaluation}

In this paper the new protocol is proposal to increase the blockchain performance. The major approach is to change the chain data structure to GraphChain and use new mining mechanism to enable parallel mining.

From the experiment, it was observed that the more miners worked, the more transactions could be processed. Although the growth was not linearly increased, adding miner to the existing blockchain can benefit the performance. It was similar to the distributed database system, when a central database running on a super computer can not meet the demand of data storage and processing capacity, people go to another direction by using thousands of small PC to build a big data system.

It's also observed that as the numbers of miners increased, the chance of all the nodes going offline would be close to zero. Parallel mining was good for building high availability blockchain systems.

There are several directions that GraphChain and parallel mining could be applied for:

1. Micro transactions. Now Bitcoin and Ethereum are valued at high price, even a small transaction fee will be expensive if converted to a legal currency. GraphChain allowing multi miners could be useful in the decentralized currency system which supports the micro transactions, as in GraphChain's design each block contain only one transaction. No block waiting time is required to confirm the transaction if enough miners are available.

2. Blockchain for data. Now privacy problems are a big issue, personal data have already become giant companies' private assets. In future there will be blockchain for certain field, for example the health data. If huge amount of data is insert into blockchain, it will be expensive to store it. GraphChain split the data unit into the smallest transaction level and the data are linked across different nodes. As a result, shard data are possible through this data structure.

3. Blockchain for banking. The banking is looking for digital currency solution. Existing blockchain can not meet the demand due to the performance and scalability issues. In this situation, new technologies will change the industry completely. The traditional banking system accepts deposits and loans to enterprises. To get deposit safely, huge costs were invested in IT systems as the election version of pen and paper. Regulation policies have been made, but they still need humans to follow it. Blockchain is a gorgeous invention, because that regulation part is no longer required for humans. The costs in regulation can be unbelievably low. A scalable and high performance blockchain will play an important role in the banking industry.

GraphChain and parallel mining are useful improvements to current blockchain systems. The performance and capacity issues will be fixed if correctly being implemented.

\section{Relatived Work}

Bitcoin-NG\cite{ref_article2}: Bitcoin-NG is proposed as an performance improvement of original Bitcoin\cite{ref_article1}. It introduces the microblock which acts a lighter block between the two original blocks on the chain. Compared with Bitcoin-NG, parallel mining is introduced to increase the system availability. A GraphChain structure is also introduced to enable miners to work together, especially when the transactions are not co-related.

BitShares\cite{ref_article3}/EOS: BitShares and EOS developed Delegated Proof-of-Stake(DPoS) from the Proof of Stake. The DPoS's delegation method tries to reduce the number of decision makers to increase the whole system's efficiency. DPoS takes a bit similar idea as leaders election: get less miners involved in transactions processing. It saves the time for decision making (less miners competing for mining) for the next block generation. GraphChain is pure PoW based, there is no PoS or DPoS consensus used by GraphChain.

IoTA/DagCoin\cite{ref_article4}: IoTA and earlier DagCoin introduce the DAG data structure, which is believed the direction for blockchain in the future. The idea Directed acyclic graph(DAG) is borrowed by lots of following systems. One of the most famous and successful examples is IoTA. In IoTA's design, there are no miners. Any transaction which is required to be confirmed by later transaction. Compare with IoTA/DagCoin, current acts more like a Bitcoin system, keeping the original PoW consensus and mining as the necessary part. Although the GraphChain borrows the DAG idea, GraphChain is not DAG.

\section*{Acknowledgment}

The authors thank Brahma OS and the team advisors' help on information collection of performance related blockchain solutions.\\

This work was supported by the National Natural Science Foundation of China under Grant No. 61701418, in part by the Jiangsu Province National Science Foundation under Grant BK20150376, in part by the Suzhou Science and Technology Development Plan under Grant SYG201516, in part by Innovation Projects of The Next Generation Internet Technology under Grant NGII20170301, in part by SIP \& XJTLU Technology Transfer Project KSF-T-03, and in part by the XJTLU research development fund projects under Grant RDF140243 and Grant RDF150246.


\begin{thebibliography}{00}



\bibitem{ref_article1}
Nakamoto, S. (2008). Bitcoin: A peer-to-peer electronic cash system.


\bibitem{ref_article2}
Eyal, I., Gencer, A. E., Sirer, E. G., \& Van Renesse, R. (2016, March). Bitcoin-NG: A Scalable Blockchain Protocol. In NSDI (pp. 45-59).

\bibitem{ref_article3}
Schuh, F., \& Larimer, D. (2015). BitShares 2.0: Financial Smart Contract Platform.

\bibitem{ref_article4}
Lerner, S. D. (2015). DagCoin: a cryptocurrency without blocks.

\bibitem{ref_article5}
Tschorsch, F., \& Scheuermann, B. (2016). Bitcoin and beyond: A technical survey on decentralized digital currencies. IEEE Communications Surveys \& Tutorials, 18(3), 2084-2123.

\bibitem{ref_article6}
Yajnik, M., Moon, S., Kurose, J., \& Towsley, D. (1999, March). Measurement and modelling of the temporal dependence in packet loss. In INFOCOM'99. Eighteenth Annual Joint Conference of the IEEE Computer and Communications Societies. Proceedings. IEEE (Vol. 1, pp. 345-352). IEEE.

\bibitem{ref_article7}
Buterin, V. (2013). Ethereum white paper. GitHub repository.

\end{thebibliography}
\end{document}